\documentclass[letterpaper]{article}
\usepackage{aaai}
\usepackage{times}
\usepackage{helvet}
\usepackage{courier}

\usepackage[pdftex]{graphicx}  
\usepackage{color}
\newcommand\blfootnote[1]{%
  \begingroup
  \renewcommand\thefootnote{}\footnote{#1}%
  \addtocounter{footnote}{-1}%
  \endgroup
}

\usepackage[ruled]{algorithm2e}
\usepackage{algorithmic}
\usepackage{amssymb}
\usepackage{amsmath}
\DeclareMathOperator*{\argmax}{arg\,max}
\usepackage{bbm}

\usepackage{booktabs} 

\begin{document}
%

\setlength\titlebox{6cm}

\title{Active Fairness in Algorithmic Decision Making}

\author{\\
\textbf{Alejandro Noriega-Campero}\footnote{Authors contributed equally to this work.}\\
MIT\\
noriega@mit.edu
\And \\
\textbf{Michiel A. Bakker}{$^*$}\\
MIT\\
bakker@mit.edu
\And \\
\textbf{Bernardo Garcia-Bulle}\\
MIT\\
bernard0@mit.edu
\And \\
\textbf{Alex Pentland}\\
MIT\\
pentland@mit.edu
}

\nocopyright
\maketitle

\begin{abstract}
Society increasingly relies on machine learning models for automated decision making. Yet, efficiency gains from automation have come paired with concern for algorithmic discrimination that can systematize inequality. Recent work has proposed optimal post-processing methods that randomize classification decisions for a fraction of individuals, in order to achieve fairness measures related to parity in errors and calibration.
These methods, however, have raised concern due to the information inefficiency, intra-group unfairness, and Pareto sub-optimality they entail. 
The present work proposes an alternative \textit{active} framework for fair classification, where, in deployment, a decision-maker adaptively acquires information according to the needs of different groups or individuals, towards balancing disparities in classification performance. We propose two such methods, where information collection is adapted to group- and individual-level needs respectively.
We show on real-world datasets that these can achieve: 1) calibration and single error parity (e.g., \textit{equal opportunity}); and 2) parity in both false positive and false negative rates (i.e., \textit{equal odds}). 
Moreover, we show that by leveraging their additional degree of freedom, \textit{active} approaches can substantially outperform randomization-based classifiers previously considered optimal, while avoiding limitations such as intra-group unfairness.
\blfootnote{$^*$Authors contributed equally to this work.}

\end{abstract}



\section{\textbf{Introduction}}

As automated decision-making systems (ADMs) have become increasingly ubiquitous\textemdash e.g., in criminal justice \cite{kleinberg2016inherent}, medical diagnosis and treatment \cite{kleinberg2015prediction}, human resource management \cite{chalfin2016productivity}, social work \cite{gillingham2015predictive}, credit \cite{huang2007credit}, and insurance \cite{siegel2013predictive}\textemdash there is widespread concern about how these can deepen social inequalities and systematize discrimination. Consequently, substantial work on defining and optimizing for algorithmic fairness has surged in the last few years. 

Inspired by domains such as race biases in criminal risk predictions \cite{flores2016false}, a substantial body of literature has focused on the problem of balancing classification errors across protected population subgroups, towards achieving equal false positive rates, false negative rates, or both (\textit{equal odds}).   
To that end, recent research has proposed ``optimal" post-processing methods that randomize decisions of a fraction of individuals to attain \textit{group fairness} \cite{hardt2016equality,pleiss2017fairness}. Yet, strong limitations of randomized approaches have been noted, such as information wastefulness, Pareto sub-optimality, and intra-group unfairness \cite{hardt2016equality,pleiss2017fairness,corbett2018measure}.

Our work aims at overcoming such limitations. We propose a complementary approach, \textit{active fairness}, where, in deployment, an ADM adaptively collects information (features) about decision subjects; gathering more information about groups or individuals harder to classify, towards achieving equity in predictive performance. Thereby, the approach leverages a natural affordance of many real-world decision systems\textemdash adaptive information collection\textemdash and allocates an ADM's information budget according to group- or individual-level needs.

\vspace{.1cm}
\noindent \textbf{Summary of contributions.} We propose two methods for achieving fairness, based on group-level and individual-level budgets. We show that, without resorting to randomization, these methods are able to achieve: a) calibration and a single-error parity constraint, and b) parity in both false positive and false negative rates (i.e., \textit{equal odds}). We show in four real-world datasets that, with constrained information budgets, active approaches can substantially outperform randomized approaches previously considered optimal (lower false positive and false negative rates). Finally, we show that classifiers using individual-level budgets in combination with active inquiry tend to dominate classifiers that use group-level budget constraints.

\vspace{.1cm}
\noindent\textbf{Intuition and motivating contexts}. Consider a patient entering a hospital seeking diagnosis, typically undergoing a progressive inquiry\textemdash measuring vitals, procuring lab tests, specialists' opinions, etc. At each step, absent sufficient certainty, the inquiry continues. Intuitively, a fair health system allocates resources to provide all patients similar-quality diagnoses. Likewise, active inquiry under cost constraints underlies contexts like disaster response, poverty mapping, homeland security, recruitment, telemedicine, refugee status determination, credit and insurance pricing, etc. 

\vspace{.1cm}
\noindent\textbf{Problem formulation.}  Let $X$ be an $n \times d$ feature  matrix. Let $X^{(q)} \subset X$ denote a query on a subset of features in $X$, with $q \subset \{0,...,d\}$, and $x_i^{(q)}$ the partial feature vector of individual $i$; and let $f(X^{(q)})$ be a predictor of class probability $P\{Y=1|X^{(q)}\}$. We study the classification context where a decision-maker can choose what information to collect about each decision subject, and seeks to maximize accuracy and fairness under an information budget constraint $\bar{b} = \frac{1}{n} \sum_{\forall i} b_i < b_{\mathit{max}}$, where $b_i = |q_i| \in [0,d]$ is the amount of information collected for individual $i$. 

Although this setting is natural to many real-world decision systems, its affordances and implications to algorithmic decision systems\textemdash at the intersection of accuracy, fairness, and cost-efficiency\textemdash have not been thoroughly studied. Here, we focus on contexts with constant costs across features. Yet we note that the active fairness framework allows generalizations to contexts with varied costs across features, as well as richer and context-specific utility functions with potential costs to decision-subjects, such as monetary, opportunity, or privacy costs.

\begin{figure*}[h!]
\centering
\includegraphics[width=.9\textwidth,trim={4cm 0cm 3.5cm 1cm},clip]{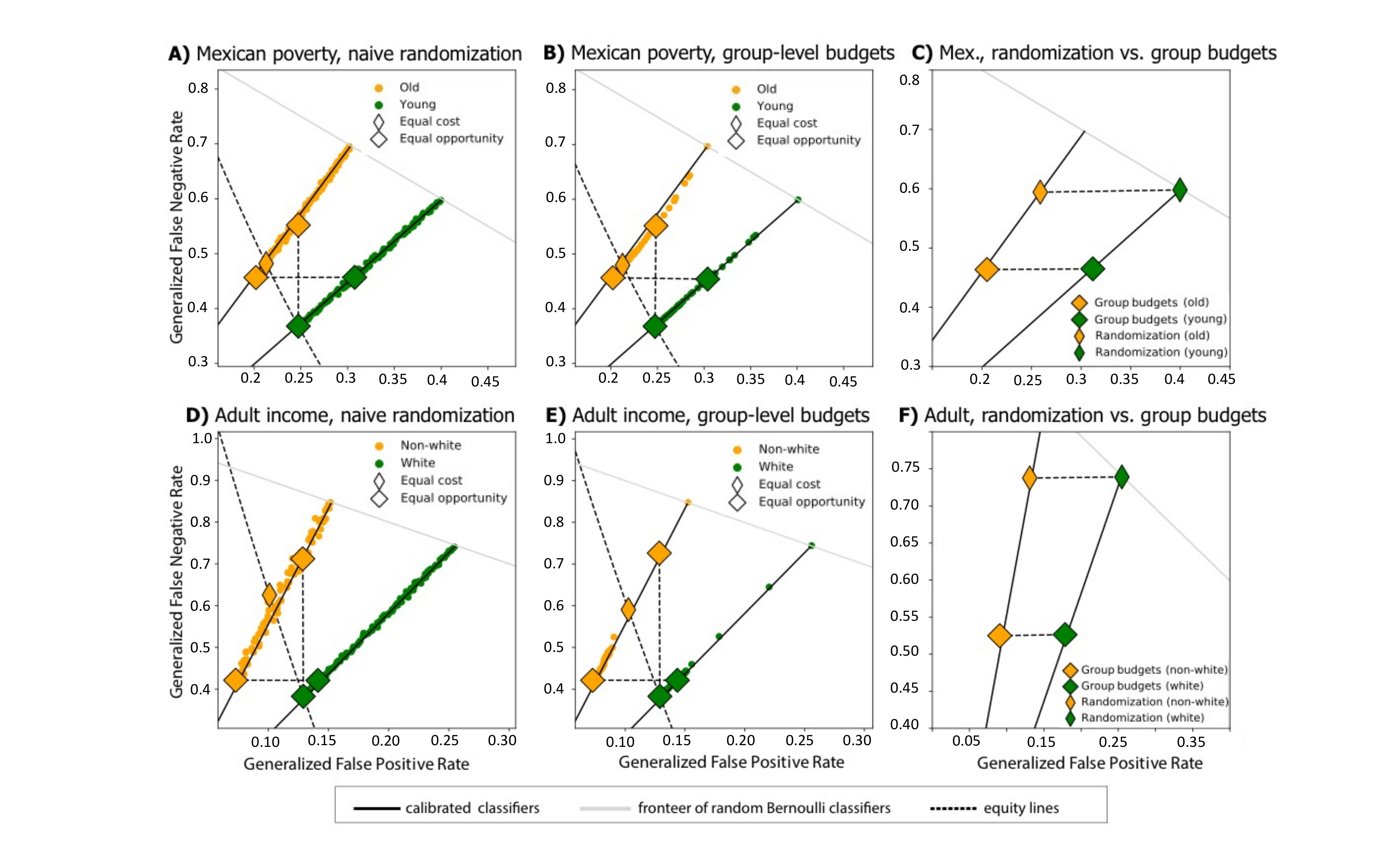}  
\vspace{-.3cm}
\caption{Achieving calibration and single error rate parity: classifiers with group-level information budgets vs. naive randomization. Rows correspond to analysis on two different datasets: the \textit{Mexican poverty} and \textit{adult income} datasets. Green and yellow colors correspond to error rates for two population subgroups (e.g., white and non-white individuals). Solid black lines represent the space of calibrated classifiers. Panels in the first column (A and D) show the generalized false positive and false negative rates (GFPR and GFNR) of classifiers that randomize an increasing proportion of individuals (0\% to 100\%), as in \cite{pleiss2017fairness}. In line with \cite{pleiss2017fairness}, naive randomization is able to achieve calibration and any single error parity constraint. 
Panels in the second column (B and E) show the same analysis for classifiers with group-level budgets. These classifiers are effective in achieving parity on either false positive or false negative rates, or \textit{equal cost}, while maintaining calibration; yet without resorting to naive randomization. Finally, panels in the third column (C and F) compare the efficiency of both methods, by showing the best classifiers that achieve \textit{equal opportunity} and calibration, under an information budget restriction $\bar{b} = \frac{1}{n} \sum_{\forall i} b_i < b_{\mathit{max}}$. Classifiers with group-level budgets Pareto-dominate randomized classifiers by a wide margin, i.e. for the same information budget, both population subgroups are better off, incurring substantially lower false positive and false negative errors. 
}

\label{fig:1}
\end{figure*}

\section{Related Work}
\label{related}
 \vspace{.1cm}%
\noindent\textbf{Active feature acquisition (AFA).} Several probabilistic and non-probabilistic methods exist for sequential feature querying under budget constraints \cite{gao2011active,liu2008tefe}, dating at least back to \cite{mackay1992information}, and applied in domains such as medical diagnosis \cite{gorry1968sequential}, customer targeting \cite{kanani2008prediction}, and image classification \cite{gao2011active}. To the best of our knowledge, this work is the first to study the implications AFA has to the algorithmic fairness literature and policy debate. Here, we use an approach based on probabilistic random forests, but more sophisticated methods can be used, for example, for dealing with domains with very high-dimensional input like medical images \cite{trapeznikov2013supervised}. 

\vspace{.1cm}
\noindent\textbf{Active learning.} Similar to the general AFA setting, this paper assumes that a fixed set of training data is used and that incremental features of a test sample can be queried. This differs from the active learning setting in which the system actively queries training examples that optimize learning e.g. by balancing exploration and exploitation or maximizing the expected model change \cite{settles2012active}. 
We foresee future work studying synergies in systems that attain fairness by actively choosing training samples using active learning while also applying AFA at test time. 

\vspace{.1cm}
\noindent\textbf{Notions of fairness.} Several notions of fairness and their corresponding formalizations have been proposed, most of which require that statistical properties hold across two or more population subgroups. \textit{Demographic} or \textit{statistical parity} requires that decision rates are independent from group membership \cite{calders2009building,zafar2015learning,louizos2015variational}, such that $P\{\hat{Y}=1|A=0\} = P\{\hat{Y}=1|A=1\} $, for the case of binary classification and a sensitive attribute $A\in\{0,1\}$. Most recent work focuses on meritocratic notions of fairness, or \textit{error rate matching} \cite{hardt2016equality,bechavod2017learning}, such as requiring population subgroups to have equal false positive rates (FPR), equal false negative rates (FNR), or both, i.e., $P\{\hat{Y}=1|A=0, Y=y\} = P\{\hat{Y}=1|A=1, Y=y\}, \hspace{4pt} y\in{0,1}$. In this work we focus on the latter set of fairness notions, although generalizations to others, such as statistical parity, are possible. Refer to \cite{vzliobaite2017measuring} for a survey on computational measures of fairness.

\vspace{.2cm}
\noindent\textbf{\textbf{Achieving Equal Opportunity and Equal Odds}}
\vspace{.1cm}
\label{sec:rel-cali}

\noindent Hardt et al. 2016 proposes parity in FNRs and/or parity in FPRs as a measure of unfair discrimination across population subgroups  \cite{hardt2016equality}. Parity in both types of error is referred to as \textit{equal odds}, and its relaxation, equality in only FPRs, is conceptualized as \textit{equal opportunity}, as in contexts of positive classification it means that subjects within the positive class have an equal probability of being correctly classified positive, regardless of  group membership.

\textit{Equal opportunity} can be achieved simply by shifting up or down the decision threshold $t_A$\textemdash where $\hat{Y} = \mathbbm{1}_{[t_A,1]}  \linebreak \big(\hspace{2pt} \hat{P}(Y=1|X, A) \hspace{2pt} \big) $\textemdash for group $A$ or $A^\complement$. Yet, doing so also directly affects FNRs, impeding achievement of \textit{equal odds}. In this context, Hardt et al. 2016 propose a classifier that balances both FPRs and FNRs, based on naive randomization of a fraction of individuals in the advantaged group; and prove conditions  under which the classifier is optimal with respect to accuracy \cite{hardt2016equality}.
%
%

Although effective in achieving \textit{equal odds}, these randomization-based results have been considered discouraging for reasons outlined below, and, as shown in Section \ref{sec:eq-odds}, are outperformed by active approaches.

\vspace{.2cm}
\noindent\textbf{\textbf{Achieving Calibration and Error Parity}}
\vspace{.1cm}

\noindent In many real-world uses of algorithms for risk estimation, it is common practice to require that predictions are \textit{calibrated}\textemdash
e.g., in recidivism \cite{flores2016false,corbett2017algorithmic}, child maltreatment hotlines \cite{gillingham2015predictive,chouldechova2018case}, and credit risk assessments \cite{huang2007credit}. A \textit{calibrated} estimator is one where, if we look at the subset of people who receive any given probability estimate $p\in[0,1]$, we find indeed a $p$ fraction of them to be positive instances of the classification problem. In the context of credit assignment, for example, we would expect a $p$ fraction of credit applicants with estimated default risk of $p$ to default. Moreover, in the context of algorithmic fairness across population groups, it is desired that calibration holds for each group \cite{flores2016false}. 

Calibration is not necessary nor sufficient to achieve parity in classification errors \cite{corbett2018measure}. However, it is particularly desirable in cases where the output of an algorithm is not directly a decision but used as input to the subsequent judgment of a human decision-maker. In such contexts, risk estimates of an uncalibrated algorithm would carry a different meaning for different groups (e.g., African-American and white defendants), and hence their use in informing human judges' decisions would likely entail disparate impact.

Recently, Kleinberg et al. 2016 demonstrated that a tension exists between minimizing error disparity across different population groups and maintaining calibrated probability estimates \cite{kleinberg2016inherent}. In particular, it showed that calibration is compatible only with a single error constraint (i.e. equal FNR or equal FPR). On the same vein, Pleiss et al. 2017 showed that the results hold for even a strong relaxation of \textit{equal odds}, named \textit{equal cost}, where FPRs and FNRs are allowed to compensate one another according to a cost function \cite{pleiss2017fairness}. Finally, they propose a method that, using naive randomization, is able to achieve parity on either error rate or \textit{equal cost}. We compare our methods to these benchmarks in Section \ref{sec:eq-op}.

\pagebreak
\noindent\textbf{\textbf{Objections to naive randomization}}
\vspace{.1cm}

\noindent The above results on achieving \textit{equal odds}, as well as on jointly achieving calibration and a single error parity measure, rely on naive randomization as means to fairness. Hence, they have been interpreted as unintuitive, discouraging, and unsettling \cite{hardt2016equality,pleiss2017fairness,corbett2018measure}. Several objections have been put forth against the use of naive randomization to achieve classification parity. Among them:

\noindent \textbf{Inefficiency}. As pointed out by \cite{hardt2016equality,pleiss2017fairness,corbett2018measure}, it is inefficient and appears unintuitive to withhold information that is already in hand, by naively randomizing the classification of a subset of individuals.

\vspace{.1cm}
\noindent \textbf{Individual unfairness}. Classifiers based on naive randomization, such as those in \cite{hardt2016equality,pleiss2017fairness,corbett2018measure}, entail intra-group unfairness. Individuals who are randomized are not necessarily those with higher uncertainty but simply the ones who were unlucky, hence breaking ordinality between the probability of classification error and the underlying uncertainty. 

\vspace{.1cm}
\noindent \textbf{Pareto sub-optimality and undesirability}. Consider an unconstrained and unfair classifier $\hat{Y}_U$, which incurs higher errors on group $A$ than group $B$; and consider an alternative "fair" classifier $\hat{Y}_F$, where a percentage of individuals of group $B$ are randomized to achieve parity in errors. Considering groups $A$ and $B$ as the system's stakeholders, we note that the original unfair classifier $\hat{Y}_U$ Pareto dominates the fair alternative $\hat{Y}_F$, i.e.: the disadvantaged group $A$ will be indifferent, as its classification remained unchanged, while group $B$ will strongly prefer $\hat{Y}_U$, the original classifier before accuracy was degraded by randomization. No group would prefer $\hat{Y}_F$.

%
%
%
%

\section{\textbf{Active Fairness}}
\label{sec:active}

The present work explores \textit{active feature acquisition} approaches for achieving fairness, where a decision-maker adaptively acquires information according to the needs of different groups or individuals, in order to balance disparities in classification performance. This section defines two such strategies, one that allocates group-level information budgets\textemdash constant for all members of a group\textemdash and one that allocates individual-level information budgets, which are computed dynamically at test time. Sections \ref{sec:eq-op} and \ref{sec:eq-odds} demonstrate their use and advantages in attaining fairness.

\vspace{.15cm}
\noindent\textbf{Preliminaries.} We denote data of each decision subject as a pair $(x,y)$, where $x$ is a feature vector of dimensionality $d$, and $y$ is an outcome of interest. Let
$S=(x^i, y^i)^{n}_{i=1}$ denote a labeled dataset, and $A \subset S$ represent a population subgroup. Let $\hat{Y}(X)$  be a binary classifier. We denote by $ \mathit{FPR}_{A}(\hat{Y})$ and $ \mathit{FNR}_{A}(\hat{Y})$ the false positive and false negative rates of $\{(x,y) \in A\}$, and define disparity measures with respect to $A$ in terms of the following FPR and FNR differences:

$$ D_{\mathit{FPR}}^{A}  = \big|  \mathit{FPR}_{A}(\hat{Y}) - \mathit{FPR}_{A^\complement}(\hat{Y}) \big|$$
$$ D_{\mathit{FNR}}^A  = \big|  \mathit{FNR}_{A}(\hat{Y}) - \mathit{FNR}_{A^\complement}(\hat{Y}) \big|$$

\textit{Equal opportunity}\textemdash or FNR parity\textemdash with respect to $A$ requires that $D_{\mathit{FNR}}^A=0$, while \textit{equal odds} requires that both $ D_{\mathit{FNR}}^A  = D_{\mathit{FPR}}^A = 0 $ \cite{hardt2016equality}.

\subsection{Group-Level Information Budgets}
\vspace{.1cm}

Let $b_A, b_B$ be the information budgets for population sub-groups $A,B$. We define predictor $h^g$ with group-level information budgets $b_A, b_B$ by:
\[
h^g(x_i) = 
     \begin{cases}
       f(x_i^{(q_A)})  &\quad\text{if}   \hspace{8pt} i \in A\\
       f(x_i^{(q_B)})  &\quad\text{if}  \hspace{8pt}  i \in B\\
     \end{cases}
\]
\noindent where $q_A, q_B$ are feature sets that satisfy $|q_A|= b_A$ and $|q_B|= b_B$. Sections \ref{sec:eq-op} and \ref{sec:eq-odds} show how decision-makers can achieve calibration and group-level equity by allocating budgets $b_A, b_B$.

\subsection{Individual-Level Information Budgets}
\label{sec:individual}

Beyond group-level budgets, an ADM may adaptively collect information of each decision subject until a confidence threshold is met, upon which a classification decision is made. Thereby, individual-level information budgets are set dynamically according to the needs of each decision subject, towards attaining equity. 

In particular, Algorithm \ref{alg:individual} specifies active inquiry at the individual level as the decision-making process that, given lower and upper probability thresholds $\alpha_l, \alpha_u \in (0, 1)$, and for each decision subject $i$, progressively expands the information set $x^{(q_i)}_{i}$ until either threshold is met, or the available feature set is exhausted. Together with the decision threshold, $\alpha_l$ and $\alpha_u$ control trade-offs between FPR and FNR. In line with related AFA methods \cite{gao2011active}, we apply early stopping to ensure we stop expanding the feature set if the classification confidence is no longer improving significantly. We estimate the parameter for early stopping using grid search while maximizing the AUC for a given budget.

We define predictor $h^\mathit{ind}$ with individual-level information budgets as $h^\mathit{ind} (x_i) = f(x_i^{(q_i)})$, where $q_i$ is the feature set according to active inquiry in Algorithm \ref{alg:individual}.

\begin{algorithm}[h]
   \SetAlgoLined
  {\bfseries Input:} data $X$, model $f$, probabilities $(\alpha_l, \alpha_u$), decision threshold $t$\;
 \For{\hspace{4pt} $i=1$ \hspace{3pt} \textbf{to} \hspace{3pt} $i=n$ \hspace{3pt}}{  
 \While{\hspace{4pt} $f(x^{(q_i)}_{i}) \leq \alpha_u \hspace{4pt} \textbf{and} \hspace{4pt} f(x^{(q_i)}_{i}) \geq \alpha_l$ \hspace{3pt} \textbf{and} \hspace{10pt}  $|q_i|<d$  \hspace{3pt} \textbf{and not}  \hspace{3pt} $e$ \hspace{3pt}}{
  $j' \leftarrow $ Get next best feature $j'\notin q_i$ \;
  $x^{(q_i)}_{i} \leftarrow x^{(q_i)}_{i} \cup x_{ij'}$   \;
  $e \leftarrow$ early\textunderscore stopping( $ x^{(q_i)}_{i} , x_{ij'}$ )   \;
 }
  $\hat{y}_i = \mathbbm{1}_{[t,1]} \big(\hspace{2pt} f(x^{(q_i)}_{i}) \hspace{2pt} \big) $ \;
  } 
  \textbf{return} \hspace{4pt} $(q_i)_{i=1}^n , \hspace{2pt} \hat{Y}$ \;
  \caption{Active inquiry at the individual level}
  \label{alg:individual}
\end{algorithm}
\vspace{-.2cm}
\subsection{Random Forest Implementation}
\label{sec:implementation-rf}

Implementation of active classifiers requires two elements: (1) a model $f$, able to estimate $P\{Y|X^{(q)}\}$ for arbitrary feature subsets $X^{(q)}$, with $q\in[0,d]$, and (2) a feature selection method for choosing expanding feature sets, either at the  group- or individual-level. 

\vspace{.1cm}
\noindent\textbf{Probabilistic model.} We implement distribution-based classification with incomplete data based on a probabilistic random forest and extending related methods for dealing with incomplete data in trees \cite{quinlan2014c4,saar2007handling}. In particular, when given an arbitrarily incomplete feature vector $x^{(q)}_i$, the algorithm traverses all possible paths of each tree according to the following rule: if value $x_{ij}$ for the current decision node is available in $q$, the search follows the path according to the node's decision function; otherwise, if the value is not available ($j\notin q$), the search follows both paths. We then compute classification probabilities as a weighted average of the leaf purity across all leaves landed on by the search. Finally, we compute the average predicted probability across all trees. Similar methods can be derived for adapting logistic regressions to admit arbitrarily incomplete feature vectors \cite{williams2005incomplete,saar2007handling}.

\vspace{.1cm}
\noindent\textbf{Static feature selection.} We first consider a static feature ranking for guiding the acquisition of additional features in Algorithm \ref{alg:individual}, based on feature importance derived from the random forest inter-trees variability. Hence, under static feature selection, given feature ranking $R$, the group-level budget classifier uses the top-$b_A$ variables in $R$ for classification of any $i\in A$, and the top-$b_B$ variables in $R$ for any $i \in B$. Similarly, the individual-level budgets classifier collects the top-$b_i$ features in $R$ in order to classify each subject $i$.

\vspace{.1cm}
\noindent\textbf{Dynamic feature selection.} In the same vein, we consider dynamic or personalized feature selection, given its potential for increased individual-level equity and overall cost-efficiency. For it we implemented a \textit{greedy} feature selection algorithm, which, for each subject $i$, and at each feature collection iteration, searches for the feature $j' \notin q_i$ that maximizes the difference between the current predicted probability $\hat{P}$ and the expected probability given that an additional feature $j'$ is queried, given by:
\begin{equation*}
j' = \argmax_{\{j:j\notin q_i, j\in[0,d]\}} \big|   \hat{P}\{y_i=1|x^{(q_i\cup j')}_i \} - \hat{P}\{y_i=1|x^{(q_i)}_i\}   \big|
\end{equation*}

\begin{figure}[h]
\centering
\includegraphics[width=.77\columnwidth]{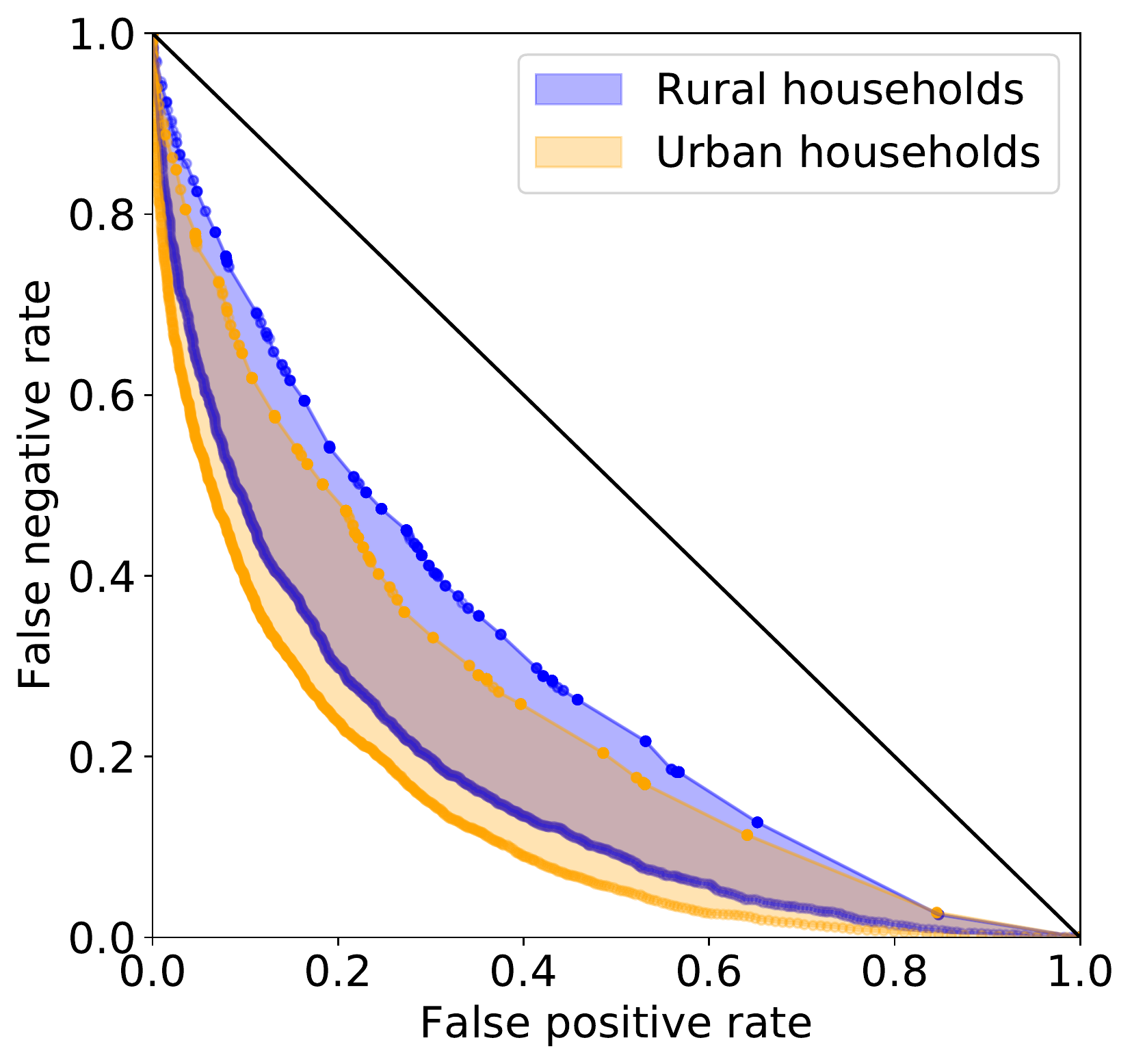} 
\vspace{-.2cm}
\caption{Achieving parity in false positives and false negatives (\textit{equal odds}) via group-level information budgets. Results correspond to the \textit{Mexican poverty dataset}. Achievable regions of classifiers for each population subgroup are plotted in blue and yellow. The outer and inner FPR-FNR curves of each achievable region correspond to classifiers using maximum and minimum information budgets. Points along the curve correspond to different values of the decision threshold. It is observed that active classifiers with group-level budgets achieve parity in both FNR and FPR (\textit{equal odds}). Moreover, they provide \textit{equal odds} solutions anywhere on the overlap of the achievable regions for both subgroups and thus along the entire FNR-FPR trade-off. 
}
\label{fig:2}
\end{figure}

\section{\textbf{Datasets}}
\label{sec:data}
We study these methods and compare them to randomization-based approaches on four real-world, public datasets. All results are computed using random $80\%/20\%$ train/test splits.

\vspace{.2cm}
\noindent\textbf{Mexican poverty}.
Targeted social programs are challenged with household poverty prediction in order to determine eligibility \cite{ibarraran2017conditional}. This dataset is extracted from the Mexican household survey 2016, which contains ground-truth household poverty levels, as well as a series of visible household features on which inferences are based. The dataset comprises a sample of 70,305 households in Mexico, with 183 categorical and continuous features, related to households' observable attributes and other socio-demographic features. Classification is binary according to the country's official poverty line, with $36\%$ of the households having the label poor. We study fairness across groups defined by a)  young and old families, split by the mean (where $53\%$ are young), and b) across families living in urban and rural areas (where $64\%$ are urban). 

\vspace{.1cm}
\noindent\textbf{Adult income}.
The Adult Dataset from UCI Machine Learning Repository \cite{lichman2013uci} comprises 14 demographic and occupational attributes for 49,000 individuals, with the goal of classifying whether a person's income is above \$50,000 ($25\%$ are above), and using ethnicity (whites v. non-whites) as sensitive attribute (where $86\%$ are white).

\vspace{.1cm}
\noindent\textbf{German credit}. 
The German Credit dataset from UCI Machine Learning Repository consists of 1000 instances, of which 70\% correspond to credit-worthy applicants and 30\% correspond to applicants to whom credit should not be extended. Each applicant is described by 24 attributes. The sensitive attribute describes whether people are below or above the mean age ($60\%$ is below). 

\vspace{.1cm}
\noindent\textbf{Heart health prediction}. 
The Heart Dataset from the UCI Machine Learning Repository contains 17 features from 906 adults. The target is to accurately predict whether or not an individual has a heart condition ($54\%$ has a heart condition). The sensitive attribute is whether people are below or above the mean age ($46\%$ is below). 

\begin{figure*}[h]
\centering
\includegraphics[width=0.99\textwidth,trim={0 0 0 0},clip]{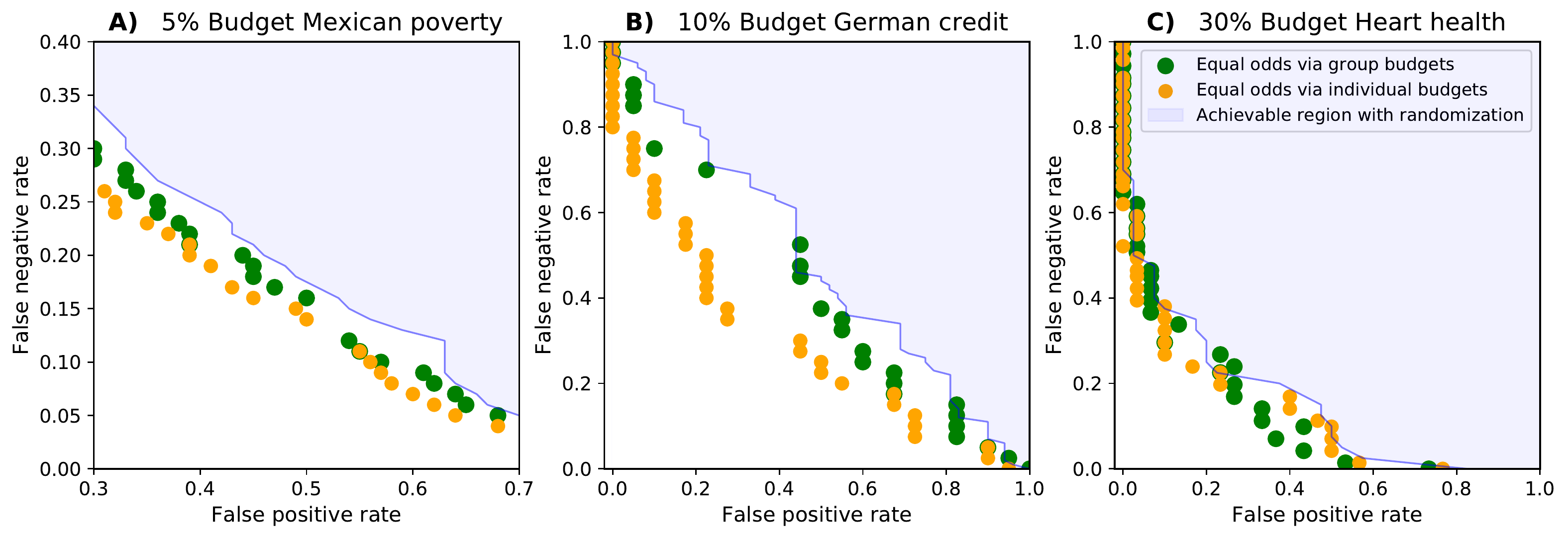} 
\vspace{-.2cm}
\caption{Achievable space of equal odds classifiers. Comparison of active classifiers with group-level (green) and individual-level (yellow) information budgets, and naive randomization (blue, following \cite{hardt2016equality}). Classifiers are constrained by an information budget $\bar{b} \leq b_{\mathit{max}}$. Figure A corresponds to \textit{Mexican poverty}, B to \textit{German credit} (center) and C to \textit{Heart health} under different budget constraints. It is observed that active classifiers yield \textit{equal odds} solutions along the FPR-FNR trade-off, without resorting to randomization. Moreover, in a budget-constraint setting, the active classifiers are substantially more information-efficient. The individual-level budget Pareto-dominates the group-level classifiers which in turn dominate randomized classifiers in budget-constrained environments.}

\label{fig:3}
\end{figure*}

\section{\textbf{Achieving Equal Opportunity \& Calibration}}
\label{sec:eq-op}

This section demonstrates how an active strategy with group-level budgets can be used to achieve calibration and single error parity, resulting in a higher efficiency and without resorting to naive randomization.

We follow \cite{pleiss2017fairness} and study predictive performance in terms of the generalized false positive (GFPR) and false negative rates (GFNR), appropriate for contexts where risk scores themselves are the outputs of the algorithm (as opposed to fully automated classification). We aim at designing classifiers that satisfy calibration and error parity. As shown by \cite{kleinberg2016inherent}, the GFNR and GFPR of all calibrated classifiers for a given group $A$ fall along the straight line with slope $(1-\mu_A)/ \mu_A$, where $\mu_A = P(Y=1| A)$ is $A$'s base-rate, and origin in the perfect classifier with $(\mathit{GFPR}, \mathit{GFNR}) = (0,0)$.

Panels A and D in Figure \ref{fig:1} show the space of calibrated classifiers achievable by naive randomization (method in \cite{pleiss2017fairness}), for the \textit{Mexican poverty} and \textit{adult income} datasets described in Section \ref{sec:data}. These replicate results from \cite{pleiss2017fairness}, showing how naive randomization of individuals in the advantaged group can, by eroding prediction performance, achieve calibration as well as either parity in false positives, parity in false negatives (but not both), or an \textit{equal cost} generalization.

Similarly, panels B and E in Figure \ref{fig:1} demonstrate how calibration and either of the three parity objectives can be achieved by adjusting information budgets according to the groups' needs, without resorting to naive randomization. Moreover, the right column in Figure \ref{fig:1} shows that classifiers with group-level budgets achieve these fairness goals with much higher efficiency in terms of information cost. In particular, we set an overall information budget restriction for both types of classifiers, equal to the minimum budget required by the naive random classifier to achieve \textit{equal opportunity}. It is observed in panels C and F that the classifiers with group-level budgets Pareto-dominate random classifiers by a wide margin, on both datasets, i.e.: for the same information budget, both population subgroups are better off, being exposed to substantially lower false positive and false negative errors.

\section{\textbf{Achieving Equal Odds}}
\label{sec:eq-odds}

This section shows how active methods with group- and individual-level information budgets can be used to achieve parity in false positives and false negatives.

Figure \ref{fig:2} illustrates the achievable regions in FPR-FNR space for classifiers with group-level information budgets, for two subgroups in the \textit{Mexican poverty dataset}. It is observed that urban households are more predictable than rural households (achievable regions closer to the origin). The yellow and purple areas comprise the achievable regions for urban and rural groups. A substantial overlap is observed, showing  a wide-ranged achievable region for \textit{equal odds}.

In a similar way, we can obtain the achievable region of active classifiers with individual-level information budgets, by varying parameters $\alpha_l < \alpha_u \in [0,1]$ of Algorithm \ref{alg:individual} (see Section \ref{sec:active}). 

We ran experiments to compare the three methods\textemdash naive randomization (as in \cite{hardt2016equality}), group-level budgets, and individual-level budgets\textemdash and their performance in achieving \textit{equal odds} solutions along the FNR-FPR trade-off. In particular, we introduce an information budget constraint $\bar{b} = \frac{1}{n} \sum_{\forall i} b_i < b_{\mathit{max}}$, and compare solutions sets that satisfy it. Solutions of individual- and group-level classifiers are discrete, due to finite sample sizes and features dimensionality.

Figure \ref{fig:3} shows results for three real-world datasets: \textit{Mexican poverty}, \textit{German credit}, and \textit{Heart health} datasets. We left out the \textit{Adult Income} dataset used in Fig.~\ref{fig:1} since there exists no overlap between achievable regions for both subgroups and therefore we cannot achieve \textit{equal odds}. Points in the FNR-FPR space were filtered to include only classifier designs that satisfied \textit{equal odds} and an overall information budget constraint $b_{\mathit{max}}$. 

It is observed that both group-level and individual-level strategies yield \textit{equal odds} solutions, covering a wide range along the FNR-FPR trade-off curve, and without resorting to naive randomization. Moreover, it is shown that both type of active classifiers are substantially more information-efficient than the randomized classifier\textemdash Pareto dominance along most of the FNR-FPR trade-off curve\textemdash leading to lower false positive and false negative errors in budget-constrained environments. Finally, active classifiers with individual-level budgets tend to dominate classifiers with group-level budgets, due to their more efficient use of information by means of personalized inquiry.

\section{\textbf{Conclusions}}
We have proposed and demonstrated methods for simultaneously achieving equal opportunity and calibration, as well as for achieving equal odds. In contrast to prior work, the \emph{active} framework does not rely on naive randomization to reach these fairness notions, avoiding several known disadvantages of randomized approaches. Instead, a decision-maker acquires partial information sets according to the needs of different groups or individuals, allocating resources equitably in order to achieve balance in predictive performance. By leveraging this additional degree of freedom, active approaches can outperform randomization-based classifiers previously considered optimal. Moreover, classifiers with individual-level budgets dominated their group-level counterparts. Finally, the extent to which the former can as well reduce intra-group unfairness is a relevant question left to future work.  

More broadly, this work illustrates how, by jointly considering information collection, inference, and decision-making processes, we can design automated decision systems that more flexibly optimize social objectives, including fairness, accuracy, efficiency, and privacy. A natural direction for future work is to consider richer utility functions relevant to real-world decision systems. We expect future studies that generalize results here presented to contexts with varying feature costs; as well as to contexts with multi-stakeholder value functions, where the opportunity, privacy, and monetary costs that inquiry and decision-making bring to decision-subjects are jointly considered as part of the adaptive inquiry process.

Lastly, a relevant path forward is to allow observations with partial feature sets both during training and test phases. The current implementation of this work necessitates access to full-feature observations at training time. More efficient training and further model refinement could be achieved under schemes that can learn from partial feature vectors, or proactively collect features at training time; allowing to incorporate a wider set of features tailored to increase prediction accuracy over different types of individuals.

\bibliographystyle{aaai}
\fontsize{9.0pt}{10.0pt} \selectfont
\bibliography{sample-bibliography}

\end{document}